\documentclass[a4paper,fleqn,usenatbib]{mnras}

\usepackage{mathptmx}

\usepackage[T1]{fontenc}
\usepackage{ae,aecompl}

\usepackage{graphicx}	
\usepackage{amsmath}	
\usepackage{amssymb}	

\title[Orbital light curve of LS\,5039]{Investigation of the energy dependence of the orbital light curve in LS\,5039}

\author[Z. CHANG et al.]{%
Z. Chang,$^{1,2}$\thanks{E-mail: changzhi@ihep.ac.cn}
S. Zhang,$^{1}$\thanks{E-mail: szhang@ihep.ac.cn}
L. Ji,$^{1,2}$
Y.P. Chen,$^{1}$
P. Kretschmar,$^{3}$
E. Kuulkers,$^{3}$
\newauthor
W.~Collmar,$^{4}$
C.Z. Liu$^{1}$
\\
$^{1}$Key Laboratory for Particle Astrophysics, Institute of High Energy Physics, Beijing 100049, China\\
$^{2}$University of Chinese Academy of Sciences, Beijing 100049, China\\
$^{3}$European Space Astronomy Centre (ESA/ESAC), Science Operations Department, Villanueva de la Ca\~nada (Madrid), Spain \\
$^{4}$Max-Planck-Institut f\"ur extraterrestrische Physik, Giessenbachstrasse, 85748 Garching, Germany
}

\date{Accepted XXX. Received YYY; in original form ZZZ}

\pubyear{2016}

\begin{document}
\label{firstpage}
\pagerange{\pageref{firstpage}--\pageref{lastpage}}
\maketitle

\begin{abstract}
LS\,5039 is so far the best studied $\gamma$-ray binary system at multi-wavelength energies.
A time resolved study of its spectral energy distribution (SED) shows that above 1\,keV
its power output is changing along its binary orbit as well as being a function of energy.
To disentangle the energy dependence of the power output as a function of orbital phase, we investigated in detail
the orbital light curves as derived with different telescopes at different energy bands.
We analysed the data from all existing \textit{INTEGRAL}/IBIS/ISGRI observations of the source and generated the most up-to-date
orbital light curves at hard X-ray energies. In the $\gamma$-ray band, we carried out orbital phase-resolved analysis
of \textit{Fermi}-LAT data between 30\,MeV and 10\,GeV in 5 different energy bands.
We found that, at $\lesssim$100\,MeV and $\gtrsim$1\,TeV the peak of the $\gamma$-ray emission
is near orbital phase 0.7,
while between $\sim$100\,MeV and $\sim$1\,GeV it moves close to orbital phase 1.0
in an orbital anti-clockwise manner. This result suggests that
the transition region in the SED at soft $\gamma$-rays (below a hundred MeV)
is related to the orbital phase interval of 0.5--1.0 but not to the one of 0.0--0.5,
when the compact object is "behind" its companion. Another interesting result is that between 3 and 20\,GeV no orbital modulation is found,
although \textit{Fermi}-LAT significantly ($\sim$18$\sigma$) detects LS\,5039.
This is consistent with the fact that at these energies, the contributions to the overall emission
from the inferior conjunction phase region (INFC, orbital phase 0.45 to 0.9) and from the
superior conjunction phase region (SUPC, orbital phase 0.9 to 0.45) are equal in strength.
At TeV energies the power output is again dominant in the INFC region
and the flux peak occurs at phase $\sim$0.7.
\end{abstract}

\begin{keywords}
gamma-rays: stars,
X-rays: individual: LS\,5039
\end{keywords}



\section{Introduction}

$\gamma$-ray binaries are a special class of X-ray binaries.
They emit high-energy $\gamma$ radiation at TeV and/or GeV energies and consist of a compact
object (neutron star or black hole) and a high-mass OB star.
LS\,5039 \citep{motch1997,aharonian2005a} is one of a few known such systems in our Galaxy.
The others are PSR\,B1259-63 \citep{johnston1992,aharonian2005b},  LS\,I\,+61$^\circ$303 \citep{albert2006,albert2009},
HESS\,J0632+057 \citep{aharonian2007} and 1FGL\,J1018.6-5856 \citep{corbet2011,abramowski2012}.
See \citet{dubus2013} for a recent review.

\begin{figure}
\centering
  \includegraphics[width=0.4\textwidth]{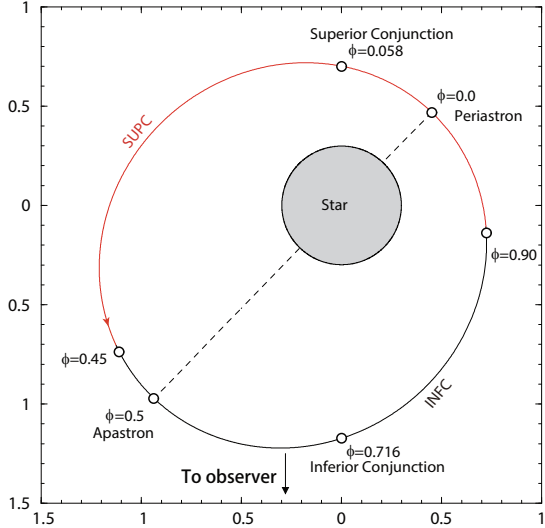}
  \caption{The LS\,5039 orbit as defined in \citet{casares2005}. The coordinates are given in units of the orbital semimajor axis.
  Relevant orbital points, such as the periastron, apastron and the conjunctions are indicated. The dashed line connects the
  periastron and apastron.
  According to the phase definitions in \citet{aharonian2006}, INFC is from $0.45<\phi<0.9$, and SUPC elsewhere with $0.9<\phi<1$ and $0<\phi<0.45$.}
  \label{fig_orbit}
\end{figure}

LS\,5039 is known as a relatively compact binary system for which the separation between the two objects is $\sim$0.1--0.2 AU.
The compact star is moving around an O6.5V main-sequence star with a period $P_{orb} \approx 3.9$ days
and a moderate eccentricity of $0.35\pm0.04$ \citep{casares2005}. A schematic view of the binary system is shown
in Figure \ref{fig_orbit}.

LS\,5039 is so far the best observed $\gamma$-ray binary at high energies.
The source was observed by \textit{Suzaku} at 1--10\,keV and 15--40\,keV, showing stable orbital light curves over a period of 6 days \citep{takahashi2009}.
At 1--10\,keV a broad peak, covering the orbital phase between 0.5 and 0.75, is seen in the orbital light curve;
the 15--40\,keV data look very similar.
At hard X-rays (25--200\,keV) the \textit{INTEGRAL} observations show a moderate orbital modulation with a peak around phase 0.75 \citep{hoffmann2009}. 
COMPTEL/\textit{CGRO} observations also find the flux to peak near phase 0.75,
with a maximal power output between 1 and 30\,MeV \citep{collmar2014}. In contrast, in the 100\,MeV -- 10\,GeV energy band \textit{Fermi}-LAT observations show that the location of the orbital flux peak is around phase 1.0 \citep{abdo2009},
while at TeV energies H.E.S.S.\ observed it to be again at the phase region around 0.7 \citep{aharonian2006}.

The SEDs at 1\,keV--1\,TeV \citep[Figure 9 of][]{collmar2014} for the inferior conjunction phase (INFC) and
the superior conjunction phase (SUPC) suggest that the relative contribution of the power output
between INFC and SUPC evolves with energy with especially a change in dominance of these two broad
regions around a few hundred MeV and a few GeV.
In order to obtain a clearer picture of the possible underlying mechanisms, especially
in these energy intervals, we have carried out a detailed analysis on the energy dependence of the
orbital light curves, using the available observational data for energies above 1\,keV.
Our analysis is based on the most up-to-date dataset of \textit{INTEGRAL}/IBIS/ISGRI between 20 and 200\,keV
(chosen among other reasons for comparison with \citet{hoffmann2009}),
and the entire \textit{Fermi}-LAT data between 30\,MeV and 10\,GeV; we use them in combination with already published \textit{Suzaku} and H.E.S.S. data.

The paper is organized with the observations and data analysis in Section 2, the results in Section 3 and finally the discussion in Section 4.

\section{Observations and data analysis}

\textit{INTEGRAL} (International Gamma-Ray Astrophysics Laboratory) was launched in 2002 October into an elliptical orbit with a period of 3 days.
The upper part of the coded-mask IBIS instrument onboard \textit{INTEGRAL},
the \textit{INTEGRAL} Soft Gamma-Ray Imager (ISGRI), consists of a Cadmium-Telluride detector array.
It is sensitive in the energy range 20\,keV -- 200\,keV.
More details about \textit{INTEGRAL}, IBIS, and ISGRI can be found in \citet{winkler2003}, \citet{ubertini2003}, and \citet{lebrun2003} respectively.

The observations of \textit{INTEGRAL} are subdivided into so-called Science Windows (ScWs), each with a typical duration of a few kiloseconds.
By selecting offset angles to the source of less than 10 degrees and exposures of larger than 100 s,
4973 public ScWs were selected for LS\,5039 in the data archive available
at the \textit{INTEGRAL} Scientific Data Center (ISDC\footnote{\url{http://www.isdc.unige.ch/}}).
The data reduction was performed by using the standard Off-line Scientific Analysis (OSA),
version 10.1, to the image step IMA.
We set {\tt OBS1$\_$PixSpread=0} to optimise the flux and signal-to-noise ratio (SNR) evaluation.

The LAT (the Large Area Telescope) onboard {\em Fermi} is an electron-positron pair production telescope operating
at energies from $\sim 20$\,MeV to $>300$\,GeV. More details about \textit{Fermi}-LAT are given in \citet{atwood2009}.

The \textit{Fermi}-LAT data reported in this paper spanned a period from 2008 August 4 (HJD 2454683.15) to 2015 March 3 (HJD 2457087.79).
These data were reduced and analysed using the \textit{Fermi} Science Tools v9r33p0 package.
Reprocessed Pass 7 data\footnote{P8 data became available after the analysis was done and a quick inspection indicates no impact on the conclusions.}
classified as source events were used.
Time intervals when the region around LS\,5039 was observed at a zenith angle less than 105$^\circ$ were selected.
The cut was used to avoid the contamination from the earth limb $\gamma$-rays,
which are produced by cosmic rays interacting with the upper atmosphere.
In our analysis, the standard pipeline, and the P7REP\_SOURCE\_V15 instrument response functions (IRFs) were adopted.
In producing the TS map, the Galactic diffuse model (gll\_iem\_v05\_rev1.fit) and the isotropic background (iso\_iem\_v05.txt),
as well as the third \textit{Fermi}/LAT catalog (3FGL, \citet{acero2015}) were properly accounted for.

\section{Results}
\subsection{Orbital phase uncertainty}

When combining data across many years, the uncertainties in the orbital phase determination can
become significant and induce phase diffusion in orbital light curves. To compare with previous
studies, we have taken as a baseline the orbital ephemeris derived by \citet{casares2005} with
$P_\mathrm{orb} = 3.90603\pm 0.00017$\,d and $T_0 = 2451943.09\pm0.10$\,(HJD). Note that \citet{hadasch2012},
based on 2.5 years of \textit{Fermi}-LAT data, independently derived a period of
$P_\mathrm{orb} = 3.90532\pm 0.0008$\,d, consistent with the previous result.

Our \textit{INTEGRAL} measurements started $\sim$767 days and ended $\sim$5018 days later than $T_0$,
which amounts to at most 0.03$\sim$0.22 uncertainty in phase due to the precision in this ephemeris (see \citet{collmar2014}).
The \textit{Fermi} data used in this analysis cover a time period from HJD 2454683.15 to HJD 2457087.79,
thus starting $\sim$2740 days and ending $\sim$5145 days after $T_0$, introducing at most $0.12\sim0.22$ in phase uncertainty.

\begin{figure}
\centering
  \includegraphics[width=0.45\textwidth]{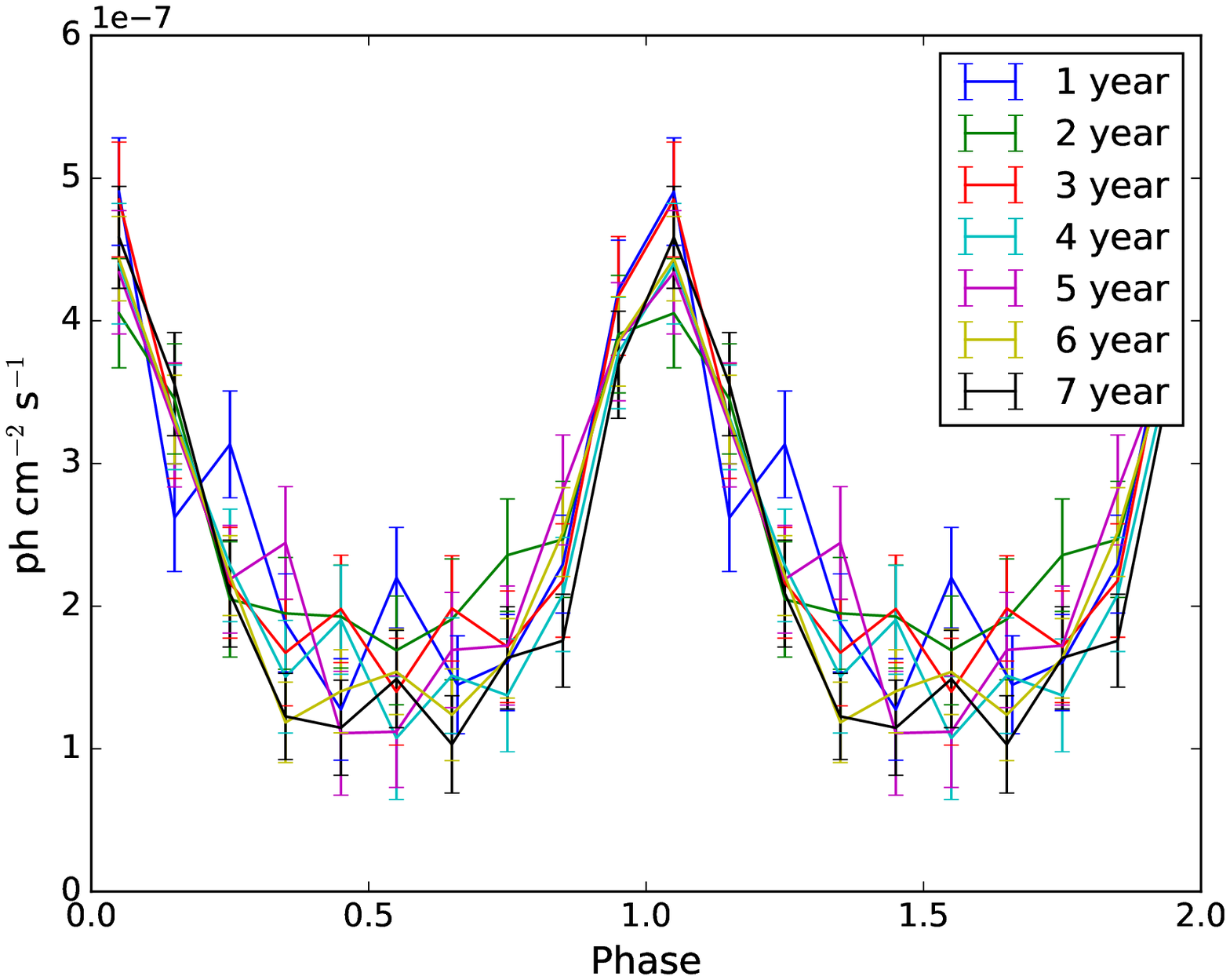}
  \includegraphics[width=0.45\textwidth]{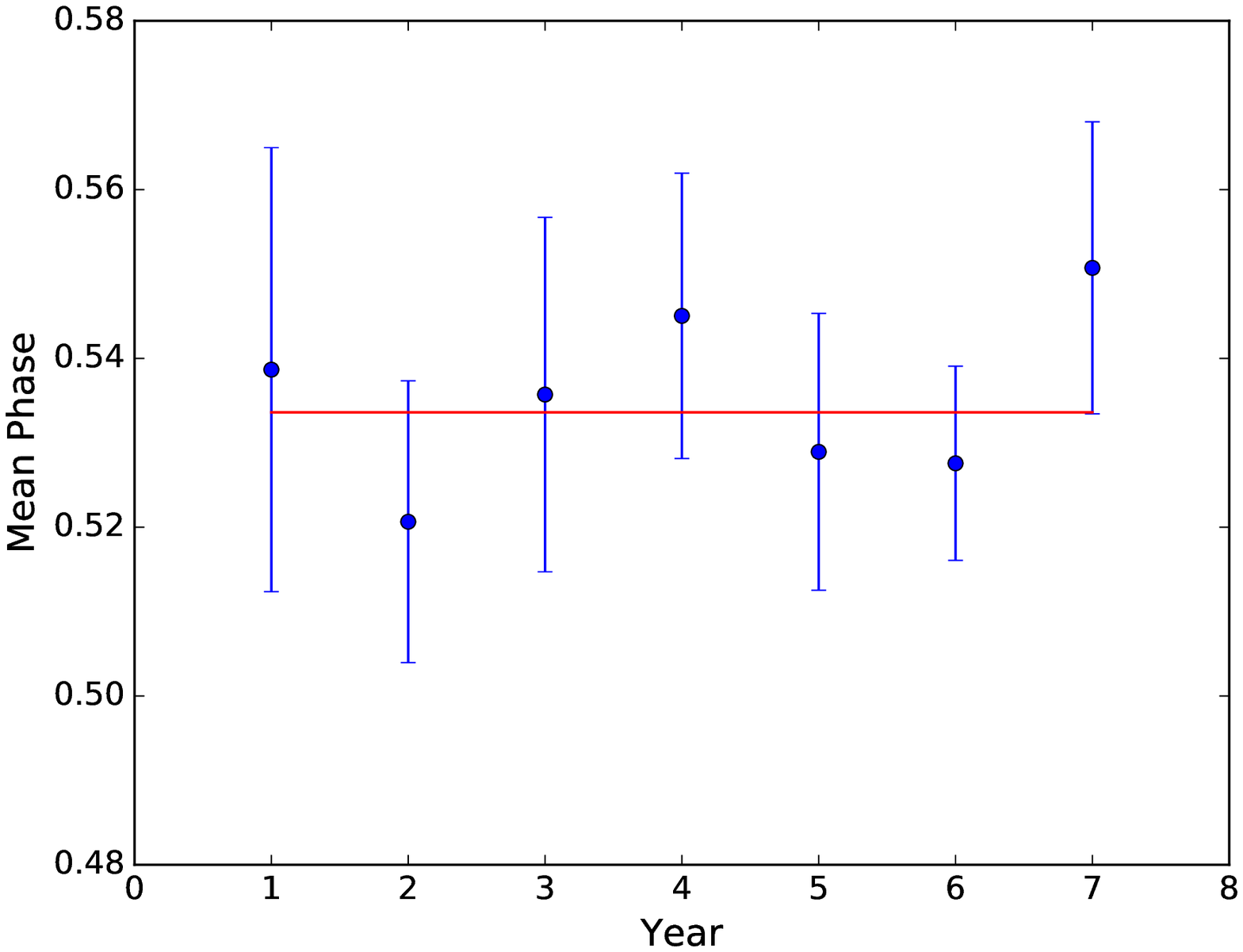}
  \caption{$Top$: \textit{Fermi}-LAT orbital light curves in the 200\,MeV -- 3\,GeV band
   for checking possible phase uncertainties caused by the inaccuracy of the ephemeris at $T_0$.
   The light curves were derived for each year of \textit{Fermi}-LAT data.
   $Bottom$: Time evolution for the phase of the minimum flux in each light curve as shown in the top panel.
   The phase value and its error were derived from an { Gaussian model} fit to the corresponding light curve.}
  \label{fig_lc_chk}
\end{figure}

In order to check for a possible phase shift we have analysed separately the
\textit{Fermi}-LAT data of each of the seven years in the 200\,MeV--3\,GeV band.
The resulting orbital light curves are shown in Figure \ref{fig_lc_chk} (top panel).

By fitting each derived light curve of the 7-year data of \textit{Fermi}-LAT with a Gaussian model,
we obtained the phase for the minimum flux in each light curve.
Figure \ref{fig_lc_chk} (bottom panel) shows the evolution of this phase value
during the time period covered by \textit{Fermi}-LAT observations.
A constant fit to these data points resulted in a reduced $\chi^2$ of 0.41 (6 degrees of freedom),
thus no significant phase shift was found.

If we alternatively take the peak flux phase values derived in the first year and the last year
of \textit{Fermi}-LAT data, we obtain an insignificant phase shift of $0.012\pm0.031$.
From this we take an upper limit of $\Delta\Phi=0.043$ as conservative error estimate, far less than the phase shift
of roughly 0.1 estimated with the error in the ephemeris as reported in \citet{casares2005}.

\begin{table}
\caption{Comparison of possible phase uncertainties for the different
      data sets as derived by using the orbital ephemeris of \citet{casares2005} (left)
      and our new derived orbital ephemeris (right). For COMPTEL, the date range was taken
      from \citet{collmar2014}, while for \textit{Suzaku} \citep{takahashi2009}, the exact times were taken
      from the \textit{Suzaku} archive. As for  H.E.S.S.\ data \citep{aharonian2006} no exact dates were
      reported, we used a date in the middle of the time interval spanned by observations.}
\begin{tabular}{l|c|c|c|c}
\hline
Telescope          & Phase Uncertainty   & Updated Uncertainty \\
\hline
\textit{Suzaku}    &  0.105              &  0.043        \\
\textit{INTEGRAL}  &  0.03--0.22         & 0.012--0.091 \\
COMPTEL            & 0.017--0.15         & 0.007--0.062 \\
\textit{Fermi}-LAT &  0.12--0.22         & 0.049--0.091 \\
H.E.S.S.           & $\sim 0.062$        & $\sim 0.026$ \\
\hline\end{tabular}
\label{tab_phase_err}
\end{table}

Using this $\Delta\Phi$ and solving the relation $\Delta\Phi= \Delta T\times\Delta P_\mathrm{orb}/P_\mathrm{orb}$
\citep{collmar2014} for the period uncertainty, we obtain a narrower uncertainty for
the orbital period $\Delta P_\mathrm{orb} = 0.00007$, compared to \citet{casares2005}.
This in turn reduces the derived phase uncertainty as demonstrated in Table \ref{tab_phase_err}.

\subsection{Skymaps and multi-wavelength light curves}
\begin{figure}
\centering
  \includegraphics[width=0.4\textwidth]{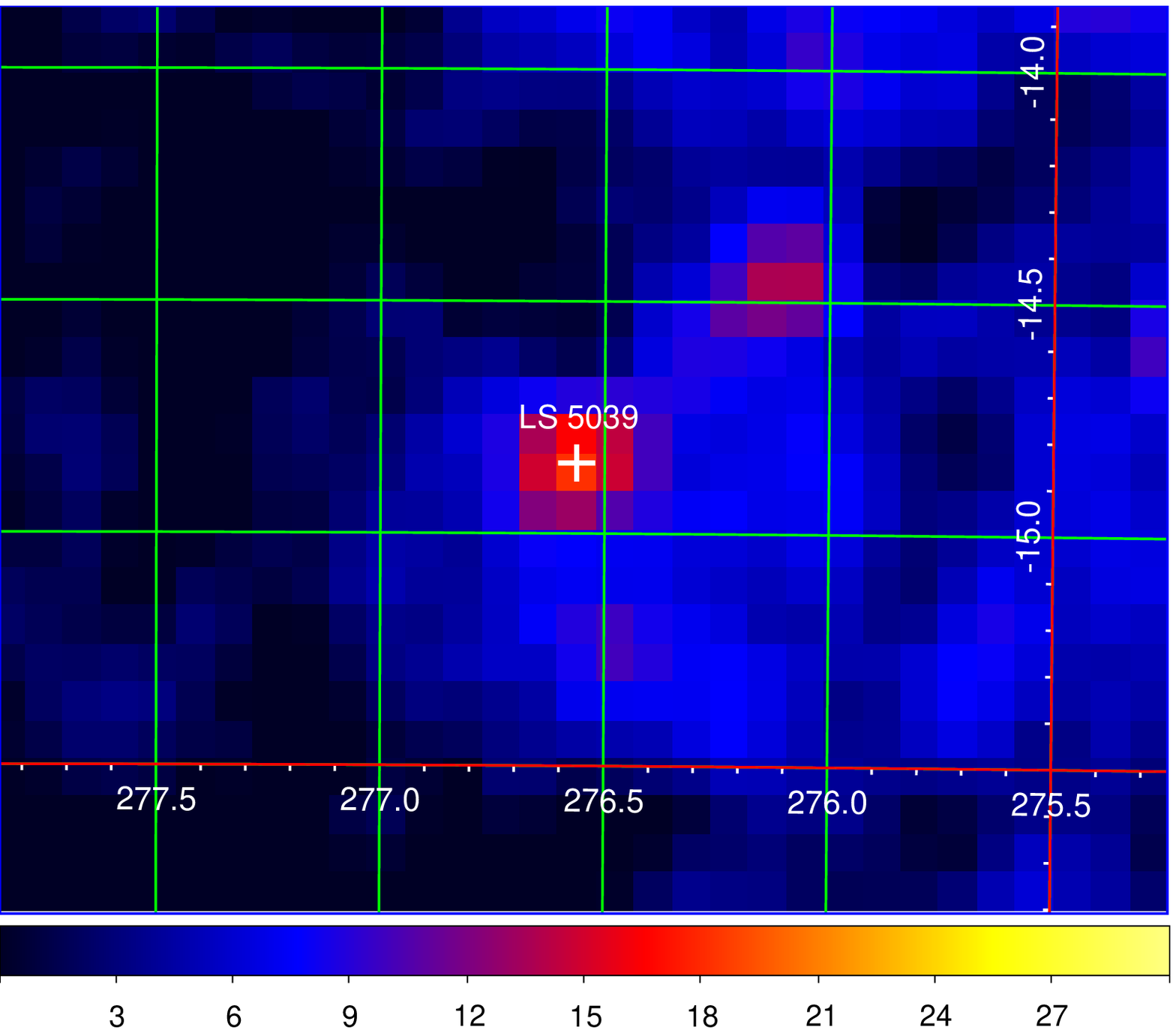}
  \includegraphics[width=0.4\textwidth]{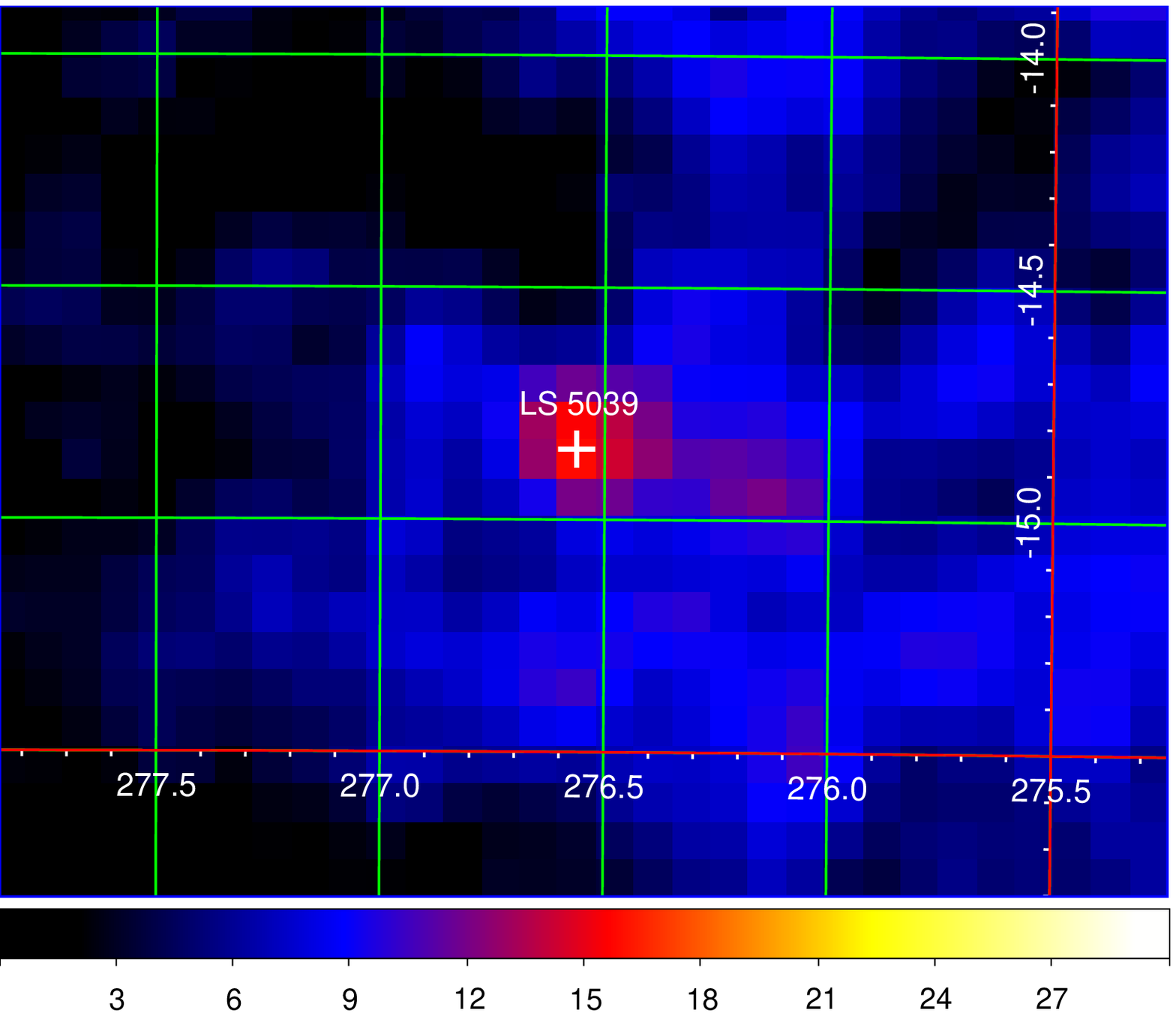}
  \caption{Detection significance maps of LS\,5039 as observed by \textit{INTEGRAL} at 20--60\,keV (top panel)
  and at 60--200\,keV (lower panel). The maps are shown in equatorial coordinates (RA, Dec).
  The source is found with detection significance values of 18.3 and 15.1 in these two energy bands, respectively.}
  \label{fig_int_sigmap}
\end{figure}

By summing over all the \textit{INTEGRAL}/IBIS/ISGRI data, we first derived the skymaps at 20--60\,keV and 60--200\,keV, respectively.
As shown in Figure \ref{fig_int_sigmap}, the source is detected with detection significance levels
of 18.3 and 15.1, respectively, based on 6.62 Ms of data,
which largely improves the detection significance 7.7 of previously reported by \citet{hoffmann2009}
based on 2.99~\mbox{Ms} of data at 25--60\,keV and using OSA version 7.
In order to produce the orbital light curves in these two energy bands, we took the source flux
extracted from the source position in individual ScWs, and then folded the flux into a light curve of 10 phase bins
using the orbital period derived above. The derived orbital light curves are shown in Figure \ref{fig_lc_all}.

\begin{figure}
\centering
  \includegraphics[width=0.35\textwidth]{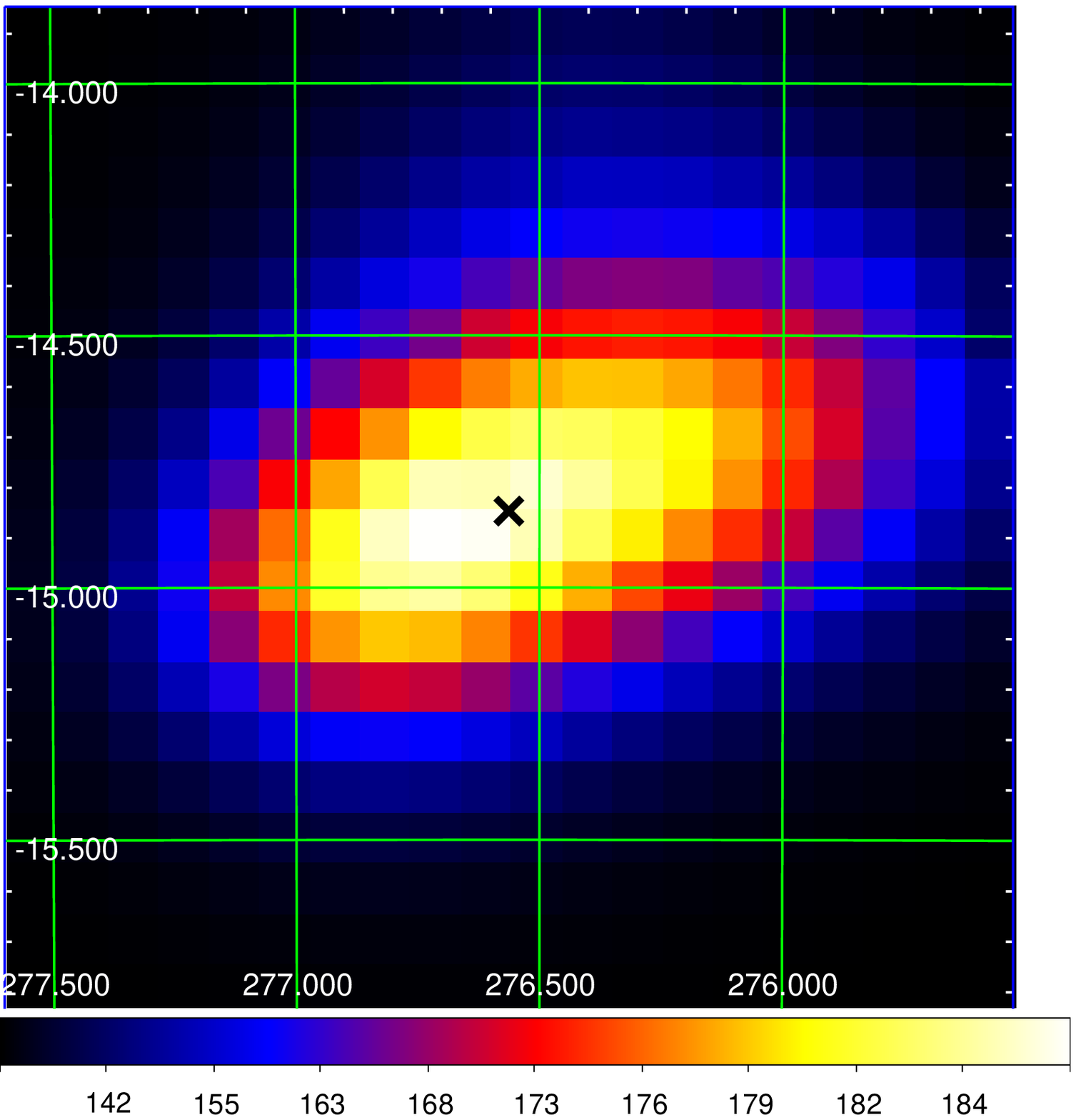}
  \includegraphics[width=0.35\textwidth]{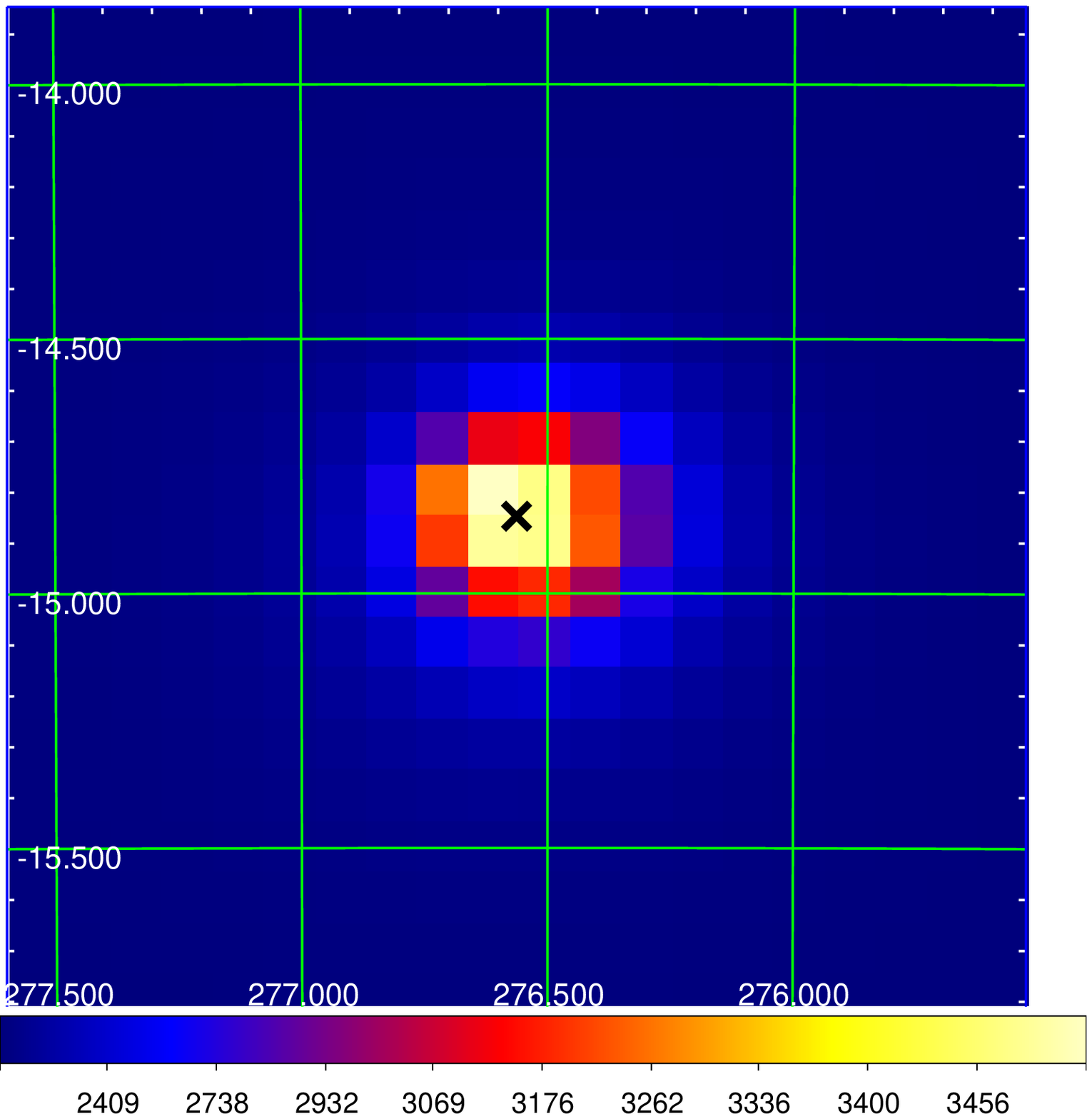}
  \includegraphics[width=0.34\textwidth]{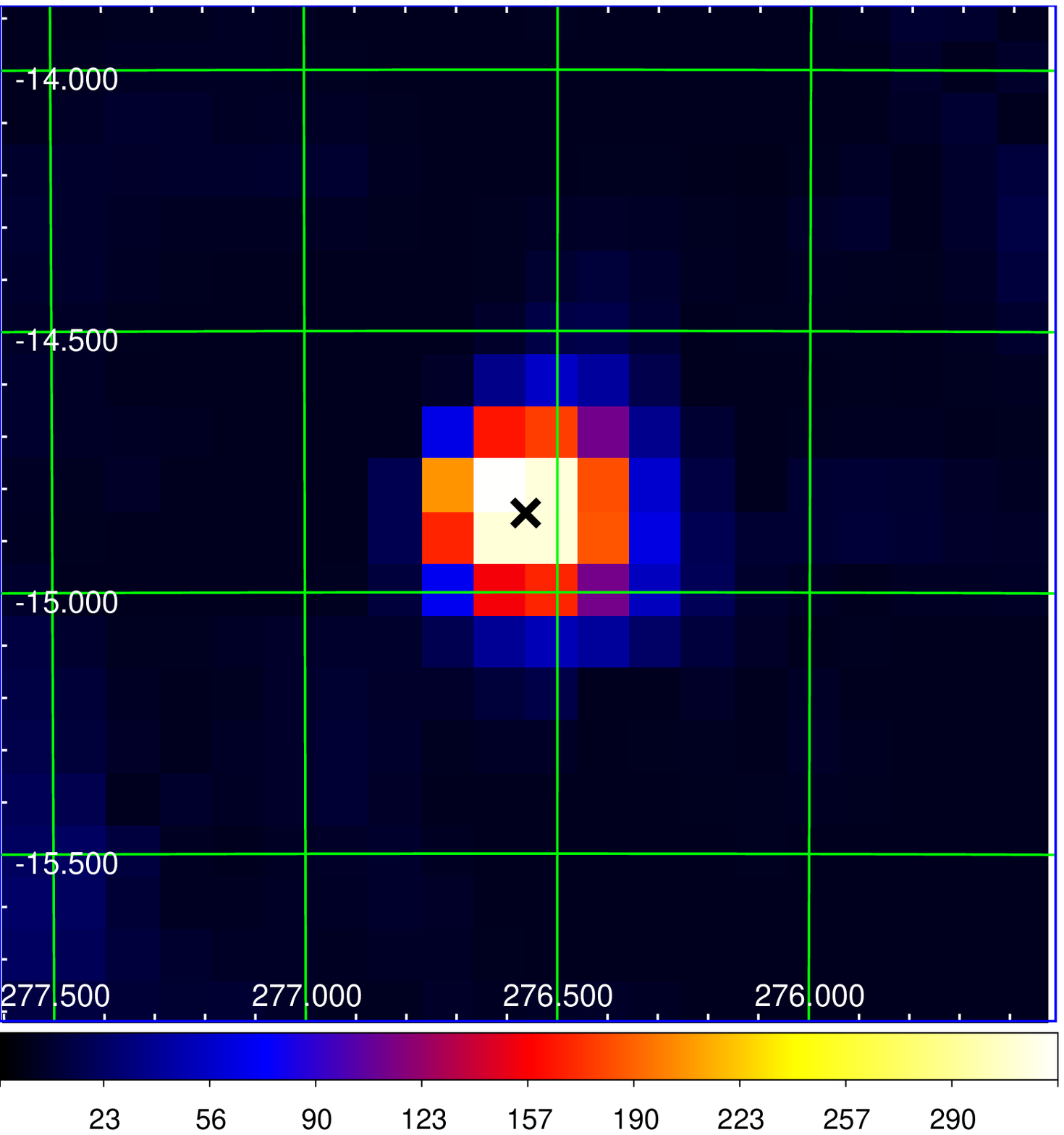}
  \caption{The TS map of LS\,5039 at 100--200\,MeV (top), 200\,MeV -- 3GeV (middle) and 3--20\,GeV (bottom)
  from the \textit{Fermi} observations.
   The source is  marked with the symbol $"\times"$ at the coordinates (RA, Dec) available in the
   \textit{Fermi}-LAT Third Source Catalog \citep{acero2015}).
   The TS value of LS\,5039 is 186, 3505 and 303, corresponding to 13.6$\sigma$, 59.2$\sigma$ and
   18.0$\sigma$ in detection significance. }
  \label{fig_ts_fermi}
\end{figure}

\begin{figure}
\centering
  \includegraphics[width=0.45\textwidth,height=1.\textwidth]{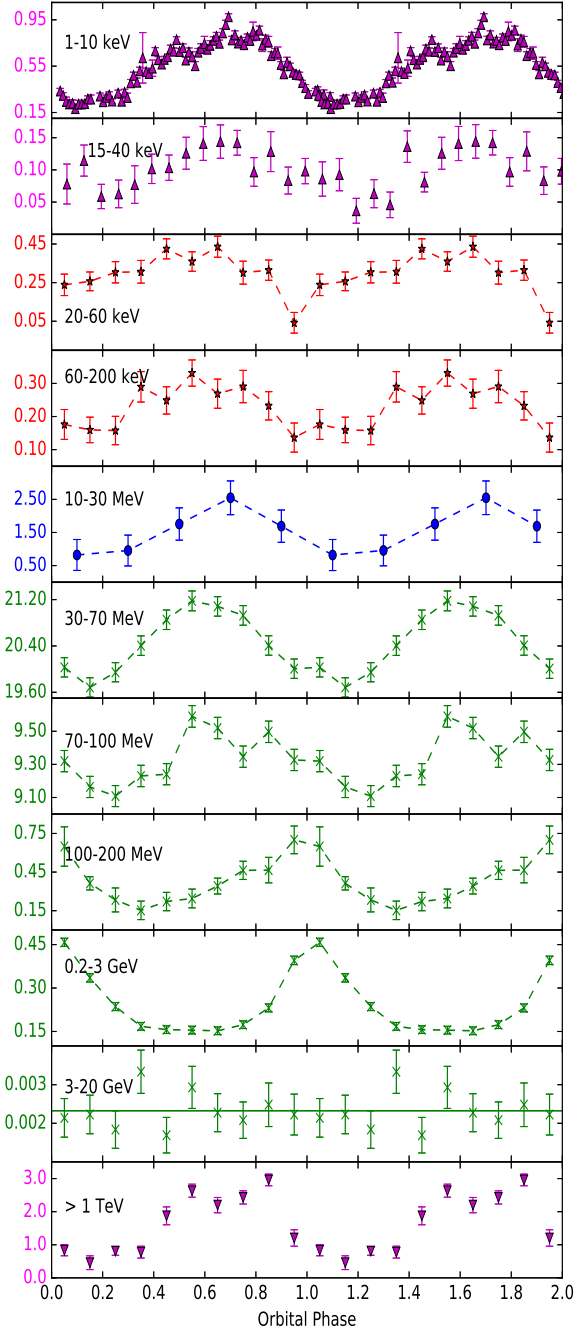}
  \caption{The top two light curves are from \textit{Suzaku} XIS and HXD, respectively, covering energy bands of 1--10\,keV and 15--40\,keV \citep{takahashi2009}.
  The next two light curves at 20--60\,keV and 60--200\,keV are from \textit{INTEGRAL}.
  The vertical axes of these four light curves are in units of counts s$^{-1}$.
  The 10--30\,MeV light curve is from COMPTEL \citep{collmar2014}, with the vertical axis in units of 10$^{-5}$ ph cm$^{-2}$ s$^{-1}$.
  The last one at $>$1\,TeV is taken from H.E.S.S.\citep{aharonian2006}, with the vertical axis in units of 10$^{-12}$ TeV$^{-1}$ cm$^{-2}$ s$^{-1}$.
  The remaining five light curves, covering the range from 30~MeV to 20 GeV, are derived in this paper with \textit{Fermi}-LAT data.
  The vertical axis is the flux in units of 10$^{-6}$ counts cm$^{-2}$ s$^{-1}$ for 30--70\,MeV and 70--100\,MeV,
  and of 10$^{-6}$ ph cm$^{-2}$ s$^{-1}$ for 100--200\,MeV, 0.2--3\,GeV and 3--20\,GeV.
  Apart from no modulation in the 3--20\,GeV band, an obvious phase shift is visible in these light curves.}
  \label{fig_lc_all}
\end{figure}

The \textit{Fermi} data were first summed into { the same} 10 phase bins using the orbital period.
The light curves were generated in five energy bands: 30--70\,MeV, 70--100\,MeV, 100--200\,MeV, 0.2--3\,GeV and 3--20\,GeV.
Limited by the \textit{Fermi} Science Tools v9r33p0 package, we could not do the likelihood analysis in both the 30--70\,MeV and 70--100\,MeV \mbox{bands}.
Instead we used the command {\tt gtexposure} to produce the light curves, assuming a power-law spectrum with a photon index of $\Gamma = 2.5$,
similar to the analysis performed in the earlier \textit{Fermi}-LAT paper \citep{abdo2009}.
The corresponding light curves are shown in Figures \ref{fig_lc_all}.
While the phase peak is around 0.5--0.75 at energies below 100\,MeV, the peak flux in orbital phase moves
toward 0.85--0.95 at 100--200\,MeV and to around 1.0 at 0.2--3.0\,GeV.
Although the TS (Test Statistic) value is about 323 at 3--20\,GeV (three TS maps of these five energy bands are shown in Figure \ref{fig_ts_fermi}),
there is no hint for an orbital modulation showing up in this light curve.

With the orbital light curves from \textit{INTEGRAL} at hard X-rays and from \textit{Fermi} at above 30\,MeV,
we were able to investigate the evolution of the orbital light curve as function of energy over a broad range.
Figure \ref{fig_lc_all} shows these new light curves together with the already published ones:
at 1--10\,keV and 15--40\,keV by \textit{Suzaku} (observed in September 2007, \citet{takahashi2009}),
1--30\,MeV by COMPTEL/CGRO (between April 1991 -- June 2000, \citet{collmar2014}), and above 1\,TeV by
H.E.S.S. (during 2004--2005, \citet{aharonian2006}).

In order to quantitatively estimate the phase shift in dif\-fer\-ent energy bands, we did a cross-correlation
analysis, using the TeV (H.E.S.S.) light curve -- where the phase of the peak flux is around 0.7 --
as reference.
We then sampled the light curves in the other energy bands, and calculated the cross-correlation
as function of phase shift for each sampled light curve with respect to the reference light curve.
The cross-correlation is described by,
\begin{center}
\begin{equation}
cor[n]=\sum^{\infty}_{k=-\infty}x[k]y[k+n]
\label{equ_cor}
\end{equation}
\end{center}
where $cor$ is the derived cross-correlation value, $x$ and $y$ are two data series.
Then we fitted the histogram of $cor$ with a Gaussian, take the mean as the shift, and the sigma as the error.
The results are shown in Table \ref{tab_phase_shift}.

\begin{table}
\caption{Results of the cross-correlation. The phase shift is given compared to the light curve of H.E.S.S..
         The negative values present a right shift, while the positive values present a left shift
         with respect to the peak in the H.E.S.S.\ orbital light curve.
         The errors only account for statistical uncertainties. }
\begin{tabular}{|l|r|r|c|c}
\hline
Telescope           & Energy        & Shift  & { Stat.Uncertainty} \\
\hline
\textit{Suzaku}     & 1--10\,keV    & 0.039  & 0.010 \\
                    & 15--40\,keV   & 0.018  & 0.083 \\
\hline
\textit{INTEGRAL}   & 20--60\,keV   & 0.178  & 0.031 \\
                    & 60--200\,keV  & 0.102  & 0.050 \\
\hline
COMPTEL             & 10--30\,MeV   & 0.001  & 0.065 \\
\hline
\textit{Fermi}-LAT  & 30--70\,MeV   & 0.077  & 0.026 \\
                    & 70--100\,MeV  & -0.037 & 0.025  \\
                    & 100--200\,MeV & -0.214 & 0.019 \\
                    & 0.2--3\,GeV   & -0.346 & 0.031 \\
\hline
H.E.S.S.            & > 1\,TeV      &  0     & 0  \\
\hline\end{tabular}
\label{tab_phase_shift}
\end{table}

Combining the results shown in Table~\ref{tab_phase_shift} and Table~\ref{tab_phase_err},
we show the derived orbital phases of the flux peaks in Figure~\ref{fig_sed_shift} (top panel), including the phase uncertainties
introduced by the orbital ephemeris.
A phase regression from $\simeq$0.6 toward $\simeq$1.0 is visible at energies above 100\,MeV.
Below about 1 MeV and above about 1 TeV the flux peak occurs at an average value of 0.68 in orbital phase.

\section{Discussion}

\subsection{Multi-wavelength variability}

\begin{figure}
\centering
  \includegraphics[width=0.45\textwidth] {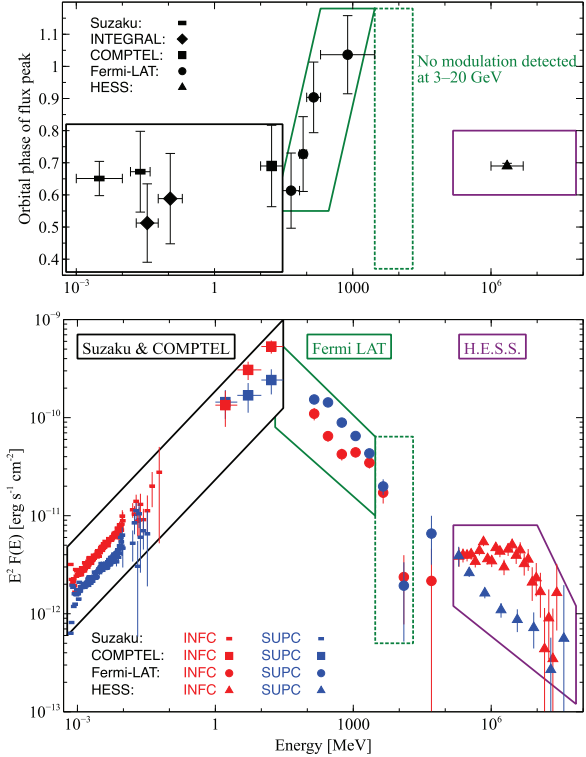}
  \caption{$Top$: The orbital phase of the flux peak as observed in different energy bands.
               Error bars include both fit uncertainties for the phase peak
               and the systematic uncertainties introduced by the uncertainty in the ephemeris.
          $Bottom$: The INFC and SUPC SEDs of LS 5039 (bottom panel, taken from \citet{collmar2014}).}
  \label{fig_sed_shift}
\end{figure}

We \textbf{have} studied the energy dependence of the flux peak in orbital phase, and
present for the first time an overall description on phase evolution across the entire high energy band.
Our results show that, at X-ray and soft $\gamma$-ray (1\,keV--30\,MeV) energies
the flux peak is located around the
region of inferior conjunction, when the compact object is in front of the massive star.
At higher energy $\gamma$-rays, from a few hundreds of MeV to about GeV, the flux peak
in orbital light curves moves from phase 0.6 toward 1.0 (from around apastron to periastron).
Generally, the orbital light curves show a single-peaked profile with a broad peak.
Interestingly, the modulation disappears at energies of a few GeV (3--20\,GeV), but shows up again
in the TeV band  with a flux peak located at around apastron to inferior conjunction, quite similar to the lowest energies.

This orbital light curve behaviour connects nicely to the high-energy SED of LS\,5039 as shown
in Figure~\ref{fig_sed_shift}. The SEDs of SUPC and INFC are subdivided into 4 boxes,
each chosen to correspond to the orbital phase evolution as a function of energy as described above.
In the 1\,keV--30\,MeV box, the power output of INFC is larger than that of SUPC,
which matches well in the orbital light curve to a broad peak around apastron.
SEDs in the 30\,MeV--3\,GeV box seem more complex due to the variability of the dominance
of the power output between SUPC and INFC.
The corresponding orbital light curves show that this may be related to the evolution
of the orbital light curve with energies: the broad flux peak is no longer located solely at
SUPC or INFC, but moves all the way in the phase from 0.6 to 1.0.
In the third box at GeV energies the two SED contributions become
equal in strength, and the light curve shows no modulation at all.
In the last box, at TeV energies, the emission from the INFC region becomes
again dominant, consistent
with an orbital light curve where the flux peaks again around apastron.

The nature of the compact object in LS\,5039, i.e., whether is is a
neutron star or a black hole is still a matter of debate.
However, it is generally thought that the system harbours a spinning neutron star,
and that the high energy emission is generated via a shock between the stellar wind of the companion and the pulsar wind of the neutron star\citep{maraschi1981}.
Support for this scenario comes from the fact that no disk/corona-like features have so far been identified
from this system: neither disk thermal emission, nor a cut-off in the energy spectrum that would
be expected from observations of a hot corona \citep{takahashi2009,torres2010}.
In a shock scenario, the kinetic power of the stellar wind and the pulsar wind is transferred to
the hot plasma which is accelerated to relativistic velocities.
High energy $\gamma$-rays are then produced via inverse Compton scattering of the shocked
plasma off the seed photons from the stellar photon field.

Up to now the high energy emission was mainly detected at either periastron
(around a few GeV) or apastron (soft X-rays, hard X-rays, soft $\gamma$-rays and $\gamma$-rays at TeV).
Therefore, emission at intermediate orbital phases is of great interest in probing further the emission model
and the physical mechanisms behind it.
The discovery of an evolution of the flux peak in the orbital phase from 0.6 at a few tens of MeV to 1.0
close to 1\,GeV provides the first clue.

\subsection{Origin of the variability}

Almost all models invoke synchrotron and inverse Compton emission to explain the multi-wavelength radiation of
systems like LS\,5039. This emission is further modified by the effects
of absorption and cooling. Along the orbit a number of conditions may vary, such as the
strength of the magnetic field, the photon field density of the massive star, the relative
direction of the line of sight and so on. Below, we discuss how these varying parameters
may impact the efficiency of both the synchrotron and the inverse Compton scattering.

As suggested in \citet{chernyakova2014}, a stellar magnetic field with the energy density in equipartition with the radiation energy density will be stronger in the vicinity of the companion star,
\begin{equation}
B_{eq}=(8\pi U_{rad})^{1/2}\simeq 1.5\times 10^2\left [\frac{T}{3\times 10^4\,\mathrm{G}}\right ]^2\left [\frac{d}{2R_*}\right ]^{-1}\,\mathrm{G}
\label{equ_magnetic}
\end{equation}
where $d$ is the distance from the massive star.
If a hot plasma population is accelerated in the stellar/pulsar wind shock,
the region closer to the compact star is expected to show synchrotron emission containing more high energy photons than regions further way.
This could account for the energy dependence of the orbital peak flux evolution.
The magnetic field around the companion star of LS\,5039 is too weak, however, to produce $\gamma$-rays via synchrotron emission.
Instead, the synchrotron emission is mostly responsible for the X-ray light curve \citep{takahashi2009}.

The modulation of the flux along the orbit can be caused by variable adiabatic cooling of electrons in the shock acceleration region \citep{takahashi2009,khangulyan2008}. The $\gamma$-ray emission
is then believed to come from inverse Compton scattering of the hot plasma accelerated by the shock off the stellar seed photons.
When the compact star moves to the region near or away from periastron,
the photons need to travel through the binary system before arriving at the observer.
The TeV photons, generated through the head-on inverse Compton scattering in the region where the seed photons have a larger density,
lose their energies through pair production \citep{dubus2013}.
Through electromagnetic cascade, more MeV and lower-energy GeV photons are generated in this region.
This can explain why the low-energy $\gamma$-ray flux peaks at the orbital phase near periastron, instead of TeV $\gamma$-rays.
Additionally, Doppler boosting can also help to minimize the TeV flux, when the compact star moves away from the observer \citep{dubus2015}.

In the region near apastron, the density of stellar photons is expected to be minimal, and
correspondingly the strength of the magnetic field in the termination shock.
In this region, absorption does not significantly affect the propagation of photons from the
shock to the observer.
In contrast to the situation near periastron, the compact star is now moving towards to the observer,
and the seed photons collide with relativistic electrons in the shock.
The head-on collision produces very high-energy photons, which can be observed easily.
The Doppler boosting also contributes to the generation of the TeV photons.

According to \citet{takahashi2009}, the particles in periastron can only be accelerated to GeV energies due to cooling and produce GeV $\gamma$-rays via inverse Compton scattering.
As the compact star is moving to periastron, the efficiency of producing synchrotron radiation is becoming larger, especially for MeV photons. 
The energy of these photons is also increased.
This is consist with our results that the phase of the flux peak in this energy band moves from phase 0.6 to 1.0.

GeV photons will be produced in INFC both by synchrotron and by inverse Compton scattering.
The tail-on scattering in the termination shock will not generate a large flux of photons that could be observed.
Though the head-on collision also takes place during the movement of the compact star from periatron
to apastron, pair production and the electromagnetic cascade limit the flux of high-energy pbhotons
that can be produced in SUPC.
This is consist with our result that the flux in the GeV band is lower than in the other energy bands.
The fact that we find no modulation in the 3--20\,GeV band, suggests that the GeV output is in balance between
the INFC region and the SUPC region.

The companion of LS\,5039 is an O-type star which is supposed to have a stable and  homogeneous wind.
It is thus generally expected that the shock structures are symmetric with respect to the
line connecting the periastron and apastron points, and, therefore, should create a double peaked
feature in the orbital light curves. This, however, is not seen in our observations.
The orbital motion as modeled in \citet{dubus2013} is not considered here, since the velocity of the compact star follows a symmetric distribution with respect to the line connecting the apastron and periastron points.
Our orbital light curve all show a single-peaked profile;
this suggests that additional facts have to be accounted for as well.
For example, the absorption through both the pair production and the electromagnetic cascade
and the influence of the opening angle of the shock wind with respect to the line of the sight, can be expected to have an effect \citep{kong2011,kong2012}.
For the latter case, the shock wave may have a smaller opening angle with respect to the line of sight in phase 0.5--1.0 than in phase 0--0.5, and hence the flux is observed to increase due to the Doppler effect \citep{dubus2010,takata2014}.

\section{Summary}
We provided new observational results on the high-energy emission of $\gamma$-ray binary LS\,5039.
By analysing all available \textit{INTEGRAL}/IBIS/ISGRI and \textit{Fermi}-LAT data up to March 2015, we provide up-to-date
orbital light curves for 7 different high energy bands.
By combining light curves in different epochs of \textit{Fermi}-LAT,
we improved the accuracy of the orbital ephemeris with respect to \citet{casares2005} by reducing its error.
We showed that the energy-dependent shape of the orbital light curves of LS\,5039
relates to the high-energy SEDs, subdivided in INFC (orbital phase 0.45 to 0.9)
and SUPC (orbital phase 0.9 to 0.45) regions.

In general, if there is a dominance of the emission from the INFC region,
we find an orbital light curve with a flux peak in the apastron to inferior conjunction region (orbital phase 0.5--0.7),
while for a SUPC-region emission dominance the flux peak in the orbital
light curves regresses to the periastron region (phase 0.9--1.0). Moreover, for radiation with energies
which are of equal importance in the SUPC and INFC regions, we find an unmodulated orbital light curve.
We discuss our findings in the framework of a neutron star pulsar model, with the assumption
that the high-energy emission is the result of a shocked region, where the stellar wind
and the pulsar wind collide. In this shocked region particles are accelerated to relativistic energies
and subsequently generate the X- and $\gamma$-radiation in the LS\,5039 system.

\section*{Acknowledgements}

We acknowledge support from the Chinese NSFC 11473027, 11133002, 11103020, XTP project XDA 04060604
and the Strategic Priority Research Program "The Emergence of Cosmological Structures"
of the Chinese Academy of Sciences, Grant No. XDB09000000. 
We thank Prof.~Tadayuki Takahashi (ISAS/JAXA) for gracefully supplying the \textit{Suzaku} data points \citep{takahashi2009}
for the SED shown in Figure~\ref{fig_sed_shift}.
We are grateful for support from the International Space Science Institute in Beijing (ISSI-BJ) for a team meeting
which motivated this publication. We thank Dr.~Jian Li and Prof.~Diego F.\ Torres for help and discussions.





\begin{thebibliography}{99}
\bibitem[Abdo et al.(2009)]{abdo2009} Abdo, A. A., et al., 2009, \apj, 706, L56
\bibitem[Abramowski et al.(2012)]{abramowski2012} Abramowski, A., et al., 2012, \aap, 541, A5
\bibitem[Acero et al.(2015)]{acero2015} Acero, F., 2015, arXiv:1501.02003v2
\bibitem[Aharonian et al.(2005a)]{aharonian2005a} Aharonian, F., et al., 2005a, \sci, 309, 746
\bibitem[Aharonian et al.(2005b)]{aharonian2005b} Aharonian, F., et al., 2005b, \aap, 442, 1
\bibitem[Aharonian et al.(2006)]{aharonian2006} Aharonian, F., et al., 2006, \aap, 460, 743
\bibitem[Aharonian et al.(2007)]{aharonian2007} Aharonian, F., et al., 2007, \aap, 469, L1
\bibitem[Albert et al.(2006)]{albert2006} Albert, J., et al., 2006, \sci, 312, 1771
\bibitem[Albert et al.(2009)]{albert2009} Albert, J., et al., 2009, \apj, 693, 303
\bibitem[Atwood et al.(2009)]{atwood2009} Atwood, W. B., et al., 2009, \apj, 697, 1071
\bibitem[Casares et al.(2005)]{casares2005} Casares, J. et al., 2005, \mnras, 364, 899
\bibitem[Chernyakova et al.(2014)]{chernyakova2014} Chernyakova et al., 2014, The X-ray Universe 2014, 237
\bibitem[Collmar \& Zhang,(2014)]{collmar2014} Collmar W.\& Zhang S., 2014, \aap, 565, A38
\bibitem[Corbet et al.(2011)]{corbet2011} Corbet, R. et al., 2011, ATel, 3221
\bibitem[Dubus et al.(2010)]{dubus2010} Dubus, G. et al., 2010, \aap, 516, 18
\bibitem[Dubus,(2013)]{dubus2013} Dubus, G., 2013, \aapr, 21, 64
\bibitem[Dubus et al.(2015)]{dubus2015} Dubus, G., 2015, \aap, 581, A27
\bibitem[Hadasch et al.(2012)]{hadasch2012} Hadasch, D. et al., 2012, \apj, 749, 54
\bibitem[Hoffmann et al.(2009)]{hoffmann2009} Hoffmann, A.D., et al., 2009, \aap, 494, L37
\bibitem[Johnston et al.(1992)]{johnston1992} Johnston, S., et al., 1992, \apj, 387, L37
\bibitem[Khangulyan  et al.(2008)]{khangulyan2008} Khangulyan, D. V., et al., 2008, Int.J.Mod.Phys.D, 17, 1909
\bibitem[Kong et al.(2011)]{kong2011} Kong, S. W., et al., 2011, \mnras, 416, 1067
\bibitem[Kong et al.(2012)]{kong2012} Kong, S. W., et al., 2012, \apj, 753, 127
\bibitem[Lebrun et al.(2003)]{lebrun2003} Lebrun, F., et al., 2003, \aap, 411, L141
\bibitem[Maraschi et al.(1981)]{maraschi1981} Maraschi, L., et al., 1981, \mnras, 194, 1P
\bibitem[Motch et al.(1997)]{motch1997} Motch, C., et al., 1997, \aap, 323, 853
\bibitem[Takata et al.(2014)]{takata2014} Takata, J., et al., 2014, \apj, 790, 18
\bibitem[Takahashi et al.(2009)]{takahashi2009} Takahashi, T. et al., 2009, \apj, 697, 592
\bibitem[Torres,(2010)]{torres2010} Torres, D. F., et al. 2010,  arXiv:1008.0483v1
\bibitem[Ubertini et al.(2003)]{ubertini2003} Ubertini, P. et al., 2003, \aap, 411, L131
\bibitem[Winkler et al.(2003)]{winkler2003} Winkler, C. et al., 2003, \aap, 411, L1
\end{thebibliography}


\end{document}